\documentclass[prd,eqsecnum,floatfix,showpacs]{revtex4}%
\usepackage{amsmath}
\usepackage{graphicx}
\usepackage{amsfonts}
\usepackage{amssymb}

\newcommand{\be}{\begin{equation}}
\newcommand{\ee}{\end{equation}}
\newcommand{\bea}{\begin{eqnarray}}
\newcommand{\eea}{\end{eqnarray}}

\begin{document}
\title{Incompatibility of Rotation Curves with Gravitational Lensing for TeVeS}

\author{Ignacio Ferreras}
\affiliation{Mullard Space Science
Laboratory, University College London, Holmbury St Mary, Dorking,
Surrey RH5 6NT, UK}

\author{Nick E. Mavromatos, Mairi Sakellariadou and Muhammad Furqaan Yusaf}
\affiliation{Department of Physics, King's College
London, Strand, London WC2R 2LS, UK}


\begin{abstract}
We constrain the one-parameter ($\alpha$) class of TeVeS models by
testing the theory against both rotation curve and strong
gravitational lensing data on galactic scales, remaining fully
relativistic in our formalism. The upshot of our analysis is that --
at least in its simplest original form, which is the only one studied
in the literature so far -- TeVeS is ruled out, in the sense that the
models cannot consistently fit simultaneously the two sets of data
without including a significant dark matter component. It is also shown
that the details of the underlying cosmological model are not relevant
for our analysis, which pertains to galactic scales. The choice of the
stellar Initial Mass Function -- which affects the estimates of
baryonic mass -- is found not to change our conclusions.
\end{abstract}


\maketitle
\section{Introduction}

With the availability of ``precision'' cosmological data, it has been
possible to formulate and test the standard ($\Lambda$CDM)
cosmological model to unprecedented accuracy, with the paradigm
consistently proving itself to be highly successful at fitting
observations \cite{teg06}. The model is based on two key components: a
framework of classical general relativity acting on a homogeneous
Friedman-Lema\^{i}tre-Robertson-Walker metric with cosmological
constant, $\Lambda~>~0$, and the presence of Cold Dark Matter
(CDM). There have been, however, other proposals for models that lie
outside the $\Lambda$CDM framework. Such models come from those who
see the as yet undetected status of dark matter and the lack of a
fundamental theory for the cosmological constant as impetus for
looking into alternative theories. In particular,
Milgrom~\cite{milgrom} put forward the MOdified Newtonian Dynamics
(MOND) to explain the flat rotation curves of galaxies without
invoking dark matter. Constructed to give flat rotation curves below
an acceleration scale, $a_0\approx 1.2 \times 10^{-10} {\rm m}/ {\rm
s}^2$, set by galactic rotation data, MOND modifies the standard
Newtonian relation between the gravitational potential and the
acceleration to $f(|{\vec a}|/a_0){\vec a}=-{\vec \nabla}\Phi_{\rm
N}$, where $\Phi_{\rm N}$ is the Newtonian gravitational field. The
function $f(x)$ is positive, smooth, and monotonic; it controls the
interpolation between the two gravitational regimes above and below
the acceleration scale. When $f(x)$ equals unity, the usual Newtonian
dynamics holds, while when $f(x)$ approximately equals its argument,
$f(x) \simeq x$, the deep MONDian regime applies.

MOND has proven itself to be very successful at fitting galactic
rotation curves, but it has been less so when applied to galaxy
clusters. A relativistic field theory counterpart to MOND has also
been expounded \cite{bek,sand}, involving the inclusion of a TEnsor,
VEctor and Scalar field (TeVeS). Since it has been constructed to
reproduce the same modification to gravity as MOND in its low
acceleration non-relativistic regime, TeVeS has not been tested using
rotation curves; it has been applied to areas such as the cosmic
microwave background \cite{sko06} and gravitational lensing, where the
relativistic aspects of the theory could be fully exploited in a way
unachievable by MOND. Most of the work on TeVeS and lensing has
remained non-relativistic \cite{zhao06,Feix,Feix2}, by simply reducing
the theory to that approximately replicated through the addition of a
scalar-field potential to the Newtonian potential. However, recent
work \cite{msy} has conducted a fully relativistic lensing analysis of
TeVeS and applied the results to a sample of six lenses from the
CASTLES survey. In order to ascertain the validity of the TeVeS/MOND
claim of explaining observations without the inclusion of dark matter,
the stellar content of these galaxies, determined by means of a
non-parametric model-independent method, was compared to the mass
predicted from lensing in the context of the TeVeS model. A
discrepancy here would be indicative of a shortcoming of the theory in
its fundamental claim. As the precise results of the comparison of
these masses depended significantly on the exact form of the MOND
$f(x)$ function, and its TeVeS equivalent $\mu(y)$, a parametrised
range of these functions was considered. The analysis showed
\cite{msy} that the parametrisation commonly used for MOND and TeVeS
requires significant quantities of dark matter, with the dependence
increasing as the parametrisation moved to the functions which best
fit rotation curves. These results suggest that a harmonisation
between fitting both rotation curves and lensing may not be possible
in these modified gravity theories.

The purpose of the present work is to fully examine the efficacy of
TeVeS, using the two distinct approaches of rotation curves and
lensing data in order to examine the validity of the implied
deficiency found in the theory~\cite{msy}. Using the two methods and
sets of data, in a fully relativistic way, we independently constrain
the degeneracy that exists within the theory regarding the precise
form of the $\mu(y)$ function. In this way we can check whether there
is any single form of the theory which can fit and describe both
rotation curves and lensing data.

The structure of the article is the following: In Section
\ref{theory}, we describe the relevant formalism, and solving the
appropriate classical equations for the metric in TeVeS.  We then use
this metric to find the modified equations for the deflection of light
and circular rotation. In Section \ref{curves} we analyse rotation
curve data for a selection of both High (HSB) and Low (LSB) Surface
Brightness galaxies. In Section \ref{lensing} we analyse gravitational
lensing for the constraints imposed by the rotation curves. We also
find the parameter which fits best the lensing data, and examine how
this fits with rotation curves. We round up with an analysis of the
results, and present our conclusions and outlook in Section
\ref{concl}.

\section{The TeVeS Model \label{theory}}

In this section we shall review the relevant formalism, leading to the
TeVeS dynamical equations, which are used when computing the
deflection angle and other quantities to be used in our fitting
procedure. We shall be brief in our analysis. For details we refer the
reader to reviews and relevant previous
work~\cite{bek,giannios,msy,sko09}.  TeVeS~\cite{bek} is a bi-metric
model in which matter and radiation do not feel the Einstein metric,
$g_{\alpha\beta}$, that appears in the canonical kinetic term of the
(effective) action, but a modified ``physical'' metric, ${\tilde
g}_{\alpha\beta}$, related to the Einstein metric by
\bea
\tilde{g}_{\alpha\beta} = e^{-2\phi}g_{\alpha\beta}-U_\alpha
U_\beta(e^{2\phi}-e^{-2\phi})~,
\eea
where $U_\mu,~\phi$ denote the TeVeS vector and scalar field,
respectively. The TeVeS action reads
\begin{eqnarray}\label{teveslagr}
S=&&\int {\rm d}^4x\ \left[\frac{1}{16\pi G}\left(R-2\Lambda\right)
  -\frac{1}{2}\{\sigma^2\left(g^{\mu\nu}-U^\mu
  U^\nu\right)\phi_{,\alpha}\phi_{,\beta}
  +\frac{1}{2}Gl^{-2}\sigma^4F(kG\sigma^2)\}\right.\nonumber\\ 
&&\ \ \ \ \ \ \ \ \ \ \left.-\frac{1}{32\pi
    G}\left\{K{\cal F}^{\alpha \beta} {\cal F}_{\alpha
    \beta}-2\lambda\left(U_\mu U^\mu + 1
  \right)\right\}\right](-g)^{1/2} +{\cal
  L}(\tilde{g}_{\mu\nu},f^\alpha,f^\alpha_{|\mu},...)(-\tilde{g})^{1/2},
\end{eqnarray}
where $k,~K$ are the coupling constants for the scalar, vector field,
respectively; $\ell$ is a free scale length related to $a_0$ (c.f
below); $\sigma$ is an additional non-dynamical scalar field; ${\cal
F}_{\mu \nu} \equiv U_{\mu,\nu} - U_{\nu,\mu}$; $\lambda$ is a
Lagrange multiplier implementing the constraint
$g^{\alpha\beta}U_{\alpha}U_{\beta}=-1$, which is completely fixed by
variation of the action; the function $F(kG\sigma)$ is chosen to give
the correct non-relativistic MONDian limit, with $G$ related to the
Newtonian gravitational constant, $G_{\rm N}$, by $G= G_{\rm N}/[
1+K/2+k/(4\pi)-2\phi_{\rm c}]$, where $\phi_{\rm c}$ is the present day
cosmological value of the scalar field. Covariant derivatives denoted
by $|$ are taken with respect to $\tilde{g}_{\mu\nu}$ and indices are
raised/lowered using the metric $g_{\mu\nu}$.

The modified equations of motion can be calculated from the
Lagrangian. For the modified Einstein equation we have~\cite{bek,giannios}
\bea
G_{\alpha\beta}+g_{\alpha \beta}\Lambda = 8\pi G\left[\tilde{T}_{\alpha\beta}
+(1-e^{-4\phi})U^{\mu}\tilde{T}_{\mu(\alpha}U_{\beta)}
+\tau_{\alpha\beta}\right]
+~\Theta_{\alpha\beta}~,
\label{metric}
\eea where \bea \tau_{\alpha\beta} &\equiv&
\sigma^2[\phi_{,\alpha}\phi_{,\beta}
  -\frac{1}{2}g^{\mu\nu}\phi_{,\mu}\phi_{,\nu}g_{\alpha\beta}
  -U^{\mu}\phi_{,\mu}(U_{(\alpha}\phi_{,\beta)}
  -\frac{1}{2}U^{\nu}\phi_{,\nu}g_{\alpha\beta})]
-\frac{G\sigma^4}{4l^2}F(kG\sigma^2)g_{\alpha\beta}~, \nonumber
\\ \Theta_{\alpha\beta} &\equiv&
K(g^{\mu\nu}U_{[\mu,\alpha]}U_{[\nu,\beta]}
-\frac{1}{4}g^{\sigma\tau}g^{\mu\nu}U_{[\sigma,\mu]}U_{[\tau,\nu]}
g_{\alpha\beta}) -\lambda U_{\alpha}U_{\beta}~.  \eea For the vector
field we obtain \bea 8\pi
G(1-e^{-4\phi})g^{\alpha\mu}U^{\beta}\tilde{T}_{\mu\beta} = K
{U^{[\alpha;\beta]}}_{;\beta}+\lambda U^{\alpha} +8\pi
G\sigma^2U^{\beta}\phi_{,\beta}g^{\alpha\gamma}\phi_{,\gamma}~,
\label{vector}
\eea
and similarly for the scalar field, namely
\bea
[\mu(y)(g^{\alpha\beta}-U^{\alpha}U^{\beta})\phi_{,\alpha}]_{;\beta} =
kG[g^{\alpha\beta}+ (1+e^{-4\phi})U^{\alpha}U^{\beta}]\tilde{T}_{\alpha\beta}~,
\label{scalar}
\eea 
with $\mu(y)$ defined by: 
\bea 
\mu(y) &=& kG\sigma^2~,\nonumber\\
 y&=&-\mu
F(\mu)-\frac{1}{2}\mu^2\frac{{\rm d}F(\mu)}{{\rm d}\mu}~,\nonumber\\ 
\quad y &=& kl^2(g^{\mu\nu}-U^{\mu}U^{\nu})\phi_{,\mu}\phi_{,\nu}~.
\label{mudef}
\eea
Motivated by the homogeneity and isotropy of the Universe, observed to
date, we assume a spherically symmetric metric 
\be
g_{\alpha\beta}{\rm d}x^\alpha {\rm d}x^\beta = -e^\nu {\rm d}t^2 +
e^\zeta({\rm d}r^2+r^2{\rm d}\theta^2 +r^2\sin^2\theta {\rm
  d}\varphi^2)~, 
\ee
where $\nu$ and $\zeta$ are both
functions of $r$. The physical metric can be written likewise, namely
\be
\tilde{g}_{\alpha\beta}{\rm d}x^\alpha {\rm d}x^\beta =
-e^{\tilde{\nu}} {\rm d}t^2 + e^{\tilde{\zeta}}({\rm d}r^2+r^2{\rm
  d}\theta^2 +r^2\sin^2\theta {\rm d}\varphi^2)~,
\label{metsys}
\ee
with the quantities $\tilde{\nu}$ and $\tilde{\zeta}$ related to
$\nu$ and $\zeta$ by
\be
\tilde{\nu} =\nu+2\phi ~~;~~
\tilde{\zeta} = \zeta-2\phi~.
\label{tildecon}
\ee
Isotropy makes the scalar field depending only on $r$, namely $\phi = \phi(r)$. 
We approximate matter as an ideal pressure-less fluid,
$\tilde{T}_{\alpha\beta}=\tilde{\rho}\tilde{u}_\alpha\tilde{u}_\beta$. By
assuming that the time-like vector field has only one non-zero temporal component, the
normalisation condition imposed by the Lagrange multiplier restricts
the vector field to be
\bea
U^{\alpha}=(e^{-\nu/2},0,0,0)~.
\label{veccon}
\eea
Considering the quasi-static case, we can take the four-velocity of the
fluid, $\tilde{u}_\alpha$, to be collinear with
$U^\alpha$, and then normalise it with respect to the
physical metric, $\tilde{g}_{\alpha\beta}$, so that $\tilde{u}_\alpha =
e^\phi U_\alpha$, leading to
\bea
\tilde{T}_{\alpha\beta} = \tilde{\rho}e^{2\phi}U_\alpha U_\beta~.
\eea
Thus, the scalar field equation, Eq.(\ref{scalar}), along with the
isotropy constraint, leads to
\bea
\frac{e^{-\frac{(\nu+3\zeta)}{2}}}{r^2}
\left[r^2\phi'e^{\frac{(\nu+\zeta)}{2}}\mu(y)\right]' = kGe^{-2\phi}\tilde{\rho}~,
\eea
where a prime denotes derivative with respect to $r$.
Upon integration, we obtain
\bea
\phi' = \frac{kGm_{\rm s}(<r)}{4\pi r^2\mu(y)}e^{-(\nu+\zeta)/2}~,
\label{scadiff}
\eea
where a scalar mass has been defined as
\bea
m_{\rm s}(<r) = 4\pi\int^r_0\tilde{\rho}
e^{\frac{\nu}{2}+\frac{3\zeta}{2}-2\phi} r^2 {\rm d}r~. \nonumber
\eea
As shown in Ref.~\cite{bek}, the scalar mass can, to a good
approximation, be equivalent to the ``proper'' mass contained in the
same volume.  Moreover, the Lagrange multiplier appearing in the
vector field, Eq.(\ref{veccon}), can be totally determined by the
vector equation, Eq.(\ref{vector}), namely
\bea
\lambda = 8\pi G(e^{-2\phi}-e^{2\phi})
\tilde{\rho}-Ke^{-\zeta}\left(\frac{\nu''}{2}
+\frac{\nu'\zeta'}{4}+\frac{\nu'}{r}\right)~.
\eea
Solving the modified Einstein equations, Eq.~(\ref{metric}), for
$\tilde{\nu}$ and $\tilde{\zeta}$ we find for the $(tt)$ and
$(\theta\theta)$ components of the stress tensor
\bea
\tau_{tt} 					
	&=& \frac{kGm_{\rm s}(<r)^2}{32\pi^2\mu(y)}
\frac{e^{-2\zeta}}{r^4}+\frac{\mu^2(y)}{4Gl^2k^2}F(\mu)e^\nu~, \nonumber \\
\tau_{\theta\theta} 	&=& -\frac{kGm_{\rm s}(<r)^2}{32\pi^2\mu(y)}
\frac{e^{-(\zeta+\nu)}}{r^2}-\frac{\mu^2(y)}{4Gl^2k^2}F(\mu)r^2e^\zeta~,
\nonumber\\
\\
\Theta_{tt} 					
&=& Ke^{\nu-\zeta}\left(\frac{(\nu')^2}{8}+\frac{\nu''}{2}
+\frac{\nu'\zeta'}{4}+\frac{\nu'}{r}\right) - 
8\pi G\tilde{\rho}(e^{-2\phi}-e^{2\phi})e^{\nu}~, \nonumber \\
\Theta_{\theta\theta} &=& \frac{K}{8}(r\nu')^2~.
\eea
Moreover, the Einstein tensor components are:
\bea
G_{tt} &=& -e^{\nu-\zeta}(\zeta''+\frac{(\zeta')^2}{4}+\frac{2\zeta'}{r})~, \nonumber \\
G_{\theta\theta} &=& \frac{r(\nu'+\zeta')}{2}+\frac{r^2(\nu')^2}{4}
+\frac{r^2(\zeta''+\nu'')}{2}~.
\eea
We thus arrive at the following system of differential equations:
\bea
\label{finalsystem}
\zeta''+\frac{(\zeta')^2}{4}+\frac{2\zeta'}{r}+e^{\zeta}\Lambda &=&
-\frac{kG^2m_{\rm
    s}^2}{4\pi\mu(y)}\frac{e^{-(\nu+\zeta)}}{r^4}  
-\frac{2\pi\mu^2(y)}{l^2k^2}F(\mu)e^\zeta\nonumber\\
&&-K
\left[\frac{(\nu')^2}{8}+\frac{\nu''}{2}+\frac{\nu'\zeta'}{4}
+\frac{\nu'}{r}\right] -
8\pi G\tilde{\rho}e^{\zeta-2\phi}~,
\nonumber\\ \frac{(\nu'+\zeta')}{2r}+\frac{(\nu')^2}{4}
+\frac{\zeta''+\nu''}{2}+e^{\zeta}\Lambda
&=& -\frac{kG^2m_{\rm s}^2}{4\pi
  \mu(y)}\frac{e^{-(\nu+\zeta)}}{r^4}-\frac{2\pi\mu^2(y)}{l^2k^2}
F(\mu)e^\zeta+\frac{K}{8}\nu'^2.
\eea
We analyse two distinct systems: lensing galaxies and galactic
rotation curves. For the former, we model the system with the spherically
symmetric  mass profile~\cite{hern}:
\bea \label{massprof}
M(<\hat{r})=\frac{M\hat{r}^2}{(\hat{r}+r_{\rm h})^2}~,
\eea
where $\hat{r}=e^{\tilde{\zeta}/2}r$ is the Schwarzschild radial
coordinate, $r_{\rm h}$ is the core radius scale, related to the
projected two-dimensional half mass radius, $R_{\rm e}$, by $R_{\rm e}
= 1.8153r_{\rm h}$~\cite{fsw05}, and $M$ is the total mass of the
galaxy. The mass, Eq.~(\ref{massprof}), specifies the density function
${\tilde \rho}$, and, as we mentioned previously, it is assumed
approximately equal~\cite{bek} to the scalar mass $m_{\rm s}(<r)$.

For the galactic rotation curves, we use a different spherically
symmetric mass profile, as outlined in Ref.~\cite{mof}, given in
Schwarzschild radial coordinates by
\bea
M(<\hat{r}) = M\left(\frac{\hat{r}}{r_{\rm c}+\hat{r}}\right)^{3\beta}~,
\eea
where $\beta = 1$ for HSB galaxies and $\beta = 2$ for LSB galaxies;
$r_{\rm c}$ is the core radius and $M$ is the total mass of the
galaxy. We remark at this point that, when converting to the physical
coordinates, a factor $e^{\tilde{\zeta}+2\phi}$ appears as the
coefficient of the cosmological constant $\Lambda$.

In order solve the equations analytically, we have used the following
approximation~\cite{bek}:
\be
e^{2\phi} \simeq e^{2\phi_{\rm c}}\left[1-\frac{kGm_{\rm s}(<r)}{2\pi r}
+\frac{k^2G^2m_{\rm s}^2(<r)}{8\pi^2r^2}+O(r^{-3})\right]~.
\ee
The precise form of the modification to gravity given in TeVeS is
largely controlled by the $\mu(y)$ function, however, since the theory
is not motivated from a microscopic theory, there exists a large
amount of freedom in choosing the form of this function. In Ref.~\cite{bek}
a toy-model function was suggested, namely
\be\label{toy}
y = \frac{3}{4}\frac{\mu^2(\mu-2)^2}{1-\mu}~.
\ee
However, it was noted in Refs.~\cite{mu1} and \cite{mu2} that when this
function is converted to its MONDian equivalent, it fits rotation
curve data worse than the standard MONDian ``simple'' form
\begin{equation}
f(x) = \frac{x}{\sqrt{1 + x^2}}~.
\label{simple}
\end{equation}
In order to improve on this point, the authors of Ref.~\cite{mu3} suggested the 
following function
\be
f(x) = \frac{2x}{1+x+\sqrt{(1-x)^2+4x}}~,
\ee
which fits better the rotation curve data in the MONDian framework.

In addition, they provided parametrised functions
$f(x)$ and $\mu(y)$ which interpolate  between the toy function, 
Eq.~(\ref{toy}), 
and the ``simple'' function, Eq.~(\ref{simple}), in both 
MOND and TeVeS frameworks, over the parameter range $0<\alpha\leq 1$:
\bea\label{muf}
f(x) = \frac{2x}{1+(2-\alpha)x+\sqrt{(1-\alpha x)^2+4x}}~; \quad
\mu(y) = \frac{\sqrt{\frac{y}{3}}}
{1-\frac{2\pi\alpha}{k}\sqrt{\frac{y}{3}}}~;~ 0<\alpha\leq 1~.
\eea
Note that, for $\alpha=0$, the function $\mu(y)$ does not coincide with the
the toy function, Eq.~(\ref{toy}), but rather approximates the
function in the high and intermediate gravity regimes. However, this
approximation is valid for our present analysis.

In fact, for our purposes, we find it convenient to use an explicit
parametrisation of $\mu$ in terms of the scalar mass $m_s$, which is
approximated by the mass profiles given directly by the data.  To this
end, we use Eq.~(\ref{scadiff}) and the definition of $y$, Eq.~(\ref{mudef}),
as well as the scalar equation of motion, Eq.~(\ref{scalar}), to write the
parametric function $\mu(y)$ in Eq.~(\ref{muf}) as:
\begin{equation}
\mu = \frac{2\pi\alpha j}{\sqrt{3}k} +
\frac{2\pi}{k}\left(\frac{\alpha^2j^2}{3}
+\frac{k^2j}{4\sqrt{3}\pi^2}\right)^{1/2}~, \quad {\rm
  where}\ ~j(r)=\left(\frac{k^{3/2}\,lG}{4\pi}\right)
\frac{m_s(<r)e^{-\tilde{\nu}/2-\tilde{\zeta}-\phi}}{r^2}~.
\label{mufinal}
\end{equation}
The corresponding $F$ function is then given by (c.f. Eq.~(\ref{mudef})):
\begin{equation}
F(\mu)=\frac{6k^3}{(4\pi\alpha)^3\mu^2}
\left[\ln\left(\frac{4\pi\alpha\mu}{k}
+1\right)^2+\frac{1}{1+\frac{4\pi\alpha\mu}{k}}
-\frac{4\pi\alpha\mu}{k}\right]~.
\label{ffunct}
\end{equation}
When $\alpha = 0$, the function $F(\mu)$ becomes singular. Since the
$\alpha = 0$ case is supposed to give a very close approximation of
the toy model function, we use for $\alpha = 0$  the explicit
expression for $F(\mu$) given in Ref.~\cite{bek}, which does not become
singular.  With all the various components given above, we are able to
numerically solve the system of differential equations to find the
functions $\tilde{\zeta}$, $\tilde{\nu}$, which specify the metric
for our choice of the $\mu(y)$ function.

\section{Fitting Rotation Curves in TeVeS\label{curves}}

In our metric system, Eq.~(\ref{metsys}), the
equation for circular orbital velocity in TeVeS reads:
\begin{equation}
V_{\rm clr}=r\frac{{\rm d}\phi}{{\rm d}t}
= \left(\frac{\tilde{\nu}'r}{2+r\tilde{\zeta'}}\right)^{1/2}~.
\end{equation}
In order to perform a fitting analysis to rotation curve data there
are two free fitting parameters: $r_{\rm c}$, which is the core radius
of the galaxy, and $\alpha$, the parameter appearing in the TeVeS
$\mu(y)$ function. For a particular choice of these parameters, we can
predict the expected rotational velocity at certain radii and this is
compared against data. Moreover, the constants in the TeVeS action,
Eq.~(\ref{teveslagr}), are taken to be
\bea
k &=& 0.01~; \quad  K = 0.01~;\quad \ell =
\sqrt{k\tilde{b}}/(4\pi\Xi a_0)~;\nonumber\\
\phi_{\rm c} &=&
0.001~; \quad  \Xi = 1+K/2-2\phi_{\rm c}~.
\eea
The values of $K$ and $k$ are constrained~\cite{bek} from solar system
tests on gravity to be $\lesssim 0.1$, and by rotation curves to be
$\gtrsim 0.001$. The scale $\ell$ is related to the MONDian
acceleration scale, $a_0$ and $\tilde{b}$. The latter quantity is
found by taking the limit of the function $y(\mu)$ when $\mu << 1$,
which then takes the form $y(\mu)\approx \tilde{b}\mu$, so for the
class of $\mu$ functions considered here, we set $\tilde{b} =
3$. Finally we note that, for $\phi_{\rm c}$, the present day value of
scalar field at cosmological scales, there are no tight constraints on
its exact value, with an approximate upper bound coming from
cosmological data.

With these constants so defined, we are able to numerically solve the
equations and perform a fitting analysis for  the galactic rotation
curves. A standard likelihood analysis is used to determine the
best fit parameters. Using $\chi^2$-statistics, the
likelihood probability distribution function for the value of $\alpha$
using data from the $i^{th}$ galaxy is defined as
\bea
L_i(\alpha) = e^{-\frac{1}{2}[\chi_i^2(\alpha)-\chi^2_{i,{\rm min}}]}~.
\eea
The combined likelihood function for all the galaxies is therefore defined
as:
\bea
L_{{\rm tot}}(\alpha) = L_1(\alpha) \times L_2(\alpha) 
\times L_3(\alpha)\times\cdots,
\eea
where the subscript on $L$ denotes the galaxy. The peak of the total
likelihood represents the most likely value of $\alpha$, and the
cumulative distribution is used to determine the 90\% confidence level
(\emph{i.e.} comparing the 5- and 95-percentile of the distribution). The
constraint on $\alpha$ from this analysis is then used to determine
lensing masses, as shown below.

In order to find the minimum value of $\chi^2$ by numerically varying
the free parameters, an adaptive grid method was implemented to give a
value of $\chi^2$ accurate up to 2 decimal places. The sample of
galaxies we chose to examine were taken from
Refs.~\cite{ver1,ver2} which explored the Tully-Fisher relation
in galaxies belonging to the Ursa Major cluster. Only those rotation
curves were examined for which there exist at least five data points
and the inner part of the velocity distribution is available,
\emph{i.e.} we excluded those galaxies for which the data only
recorded the velocity in the flat part of the rotation curve and
missed the inner sections. This gave us a sample of 6 galaxies, 2 LSB
(NGC 3972, NGC 4085) and 4 HSB (UGC 6399, UGC 6917, UGC 6923, UGC
7089). The photometric details of the galaxies are given in
Table~I.

\begin{table*}
  \begin{tabular}{l|cccc|rrr|r}
    \label{tab:photmass}
    \scriptsize{\textbf{Galaxy}} & $B-R$ & $R-I$ & $I-K$ & M$_K$ &
    M$_{\rm gas}$ & M$_{\rm st}$(Ch) & M$_{\rm st}$(Salp) & M$_{\rm N}$\\

    & \multicolumn{4}{c}{Vegamag} & \multicolumn{3}{c}{$\times 10^8$M$_\odot$} & 
    $\times 10^{10}$M$_\odot$\\ \hline
\scriptsize{NGC 3972} & $0.97$ & $0.43$ & $1.60$ & $-22.08$ & $18.07$ & $57.15$ &  $80.17$ & $3.77$\\
\scriptsize{NGC 4085} & $0.99$ & $0.46$ & $1.70$ & $-22.27$ & $15.89$ & $72.09$ & $100.93$ & $2.75$\\
\scriptsize{UGC 6399} & $0.88$ & $0.33$ & $1.56$ & $-20.33$ & $11.43$ & $11.30$ &  $15.85$ & $1.46$\\
\scriptsize{UGC 6917} & $0.86$ & $0.35$ & $1.26$ & $-21.10$ & $28.52$ & $27.61$ &  $38.37$ & $3.10$\\
\scriptsize{UGC 6923} & $0.83$ & $0.53$ & $1.16$ & $-20.36$ & $11.65$ & $13.96$ &  $19.41$ & $0.84$\\
\scriptsize{UGC 7089} & $0.85$ & $0.30$ & $1.04$ & $-20.30$ & $18.51$ & $13.77$ &  $19.14$ & $1.37$\\
\hline
  \end{tabular}
  \caption{Photometric and baryonic mass data for the galaxies in our
    sample. The photometric data are corrected for internal extinction
    and are taken from from Table~2 of Ref.~\cite{ver1}.  $M_{{\rm
        gas}}$ is the gas mass from 21~cm fluxes corrected to include
    Helium, $M_{{\rm st}}$ is the calculated mass of the stellar
    content (given for both choices of the Initial Mass Function).
    M$_{\rm N}$ gives the Newtonian estimate for the total mass, by
    using the maximum observed radial position and the maximum
    rotation velocity: M$_{\rm N}=v_{\rm MAX}^2R/G$.}
\end{table*}

The photometric details given in Table~I are used to calculate the
total baryonic content of the galaxies, fixing the mass parameter for
TeVeS as M$_{\rm gas}+$M$_{\rm st}$.  For the gas mass, we assume an
infinitely thin disk and we use the observed 21~cm line fluxes (listed
in Ref.~\cite{ver1}), using the standard translation between 21~cm flux
and neutral hydrogen mass (see, \emph{e.g.} Ref.~\cite{mof}). A correction
factor of 4/3 is necessary to take proper account of the presence of
Helium in the gas phase.

For the stellar content of the galaxies it is important to choose
realistic populations. Population synthesis models
(\emph{e.g.} Ref.~\cite{bc03}) combine our knowledge of stellar evolution
with libraries of stellar spectra to obtain simple stellar populations
(SSPs), which correspond to a family of stars all formed at the same
time, with the same \emph{metallicity}. Although SSPs can be used to describe
the spectral energy distribution of globular clusters -- which form
over very short times compared to stellar evolution timescales -- a
galaxy has a more complex distribution of stellar ages.  We follow
here the same approach as in Ref.~\cite{fsw05}, and run a grid of
models with an exponentially decaying star formation history. These
models replace the single age of a SSP by a linear superposition of
SSPs according to an exponentially decaying function of time. Hence, a
model is described by the following three parameters: (i) the
\emph{epoch} when the galaxy is \emph{born} (chosen between redshifts
z$_{\rm F}=1$ and $10$), (ii) the \emph{timescale} of the exponential
(between $\log\tau$(Gyr)$=-1$ and $+1$) and (iii) the
\emph{metallicity} (between 1/30th and twice the solar abundance). Our
``basis'' SSP models are taken from Ref.~\cite{cb07}). We run a grid
of $32\times 32\times 32$ models with a uniform distribution of the
three parameters that describe the star formation history. For each
choice, the models give a spectral energy distribution which is folded
with the response function of the standard filters $B$, $R$, $I$ and
$K$, which extend over a wide spectral range, from optical blue light
($\lambda\sim 0.4\mu$m) to near infrared ($\lambda\sim 2.2\mu$m). Our
photometric system is referenced with respect to Vega. The resulting
colours $B-R$, $R-I$ and $I-K$ are compared with the observations,
defining a likelihood which is used to determine the stellar masses by
comparing the best fit absolute magnitude in the near infrared band
($K$, $\lambda\sim 2.2\mu$m).  At longer wavelengths, stellar mass
estimates are more robust, since they minimise the attenuation from
dust and the flux mostly originates from intermediate/low mass stars,
which dominate the stellar mass budget.

In addition to age and \emph{metallicity}, the stellar mass of a population
depends sensitively on the mass distribution of stars at birth,
\emph{i.e}. the Initial Mass Function (IMF). It is generally assumed
that the IMF is universal, although its shape at the low mass end is
uncertain given how little low-mass stars contribute to the total
luminosity of a population.  In this paper we consider both the
Chabrier IMF~\cite{chab03} and the simple power law of
Salpeter~\cite{salp}. Both IMFs roughly agree at the massive end --
where observations can set significant constraints -- however, at the
low mass end, the Salpeter IMF simply extrapolates the distribution as
a power law. Recent dynamical and lensing studies (see,
\emph{e.g.} Ref.~\cite{cap,FSans}) suggest that the Salpeter IMF gives
unrealistically high stellar masses, and functions such as the
Chabrier IMF -- which replace the power law by a log-normal
distribution at low masses-- are favoured. However, given that all
those analyses were based on a ``standard'' General Relativity (GR) /
Newtonian theory, we present in this paper both choices of IMF (see,
also Ref.~\cite{sand08}).

\begin{figure}
\begin{center}
\includegraphics[width=12.5cm]{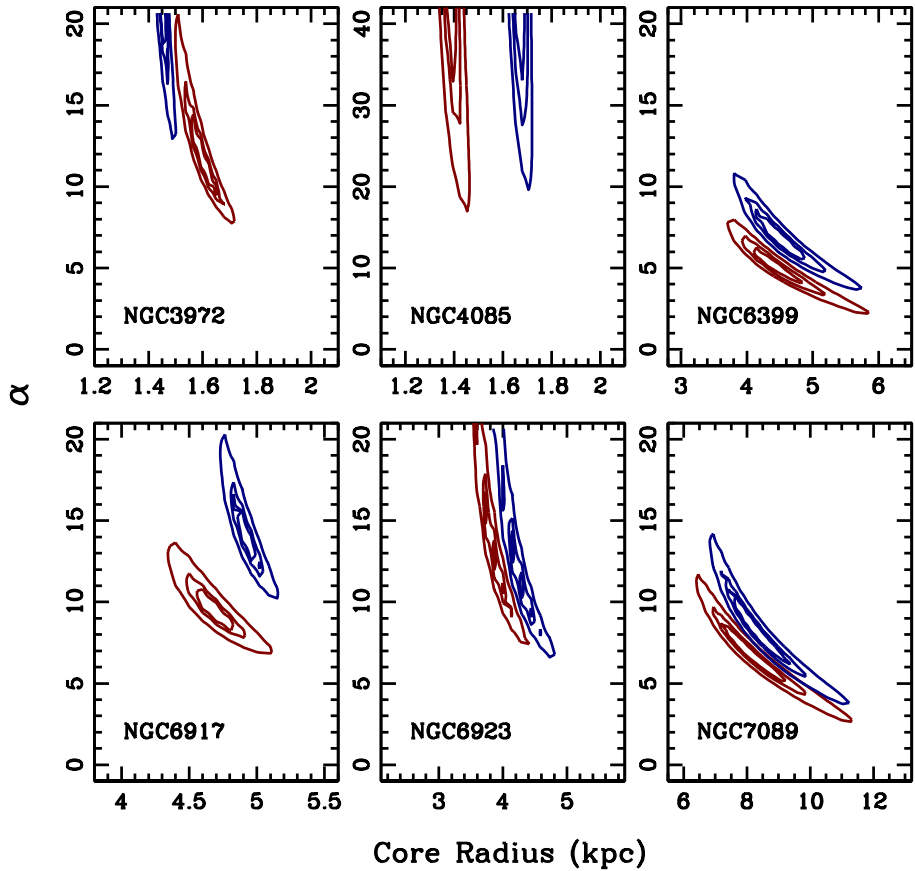}
\end{center}
\caption{Likelihood maps of the best fit for the rotation curves of
  disk galaxies, treating the core radius (horizontal axis) and the
  TeVeS $\alpha$ parameter (vertical axis) as free parameters. The red
  (blue) contours correspond to the 75\%, 90\% and 95\% confidence
  levels of the Chabrier (Salpeter) Initial Mass Function,
  respectively.}
\label{fig:chi2maps}
\end{figure}

\begin{figure}[h]
\begin{center}
\includegraphics[width=12.5cm]{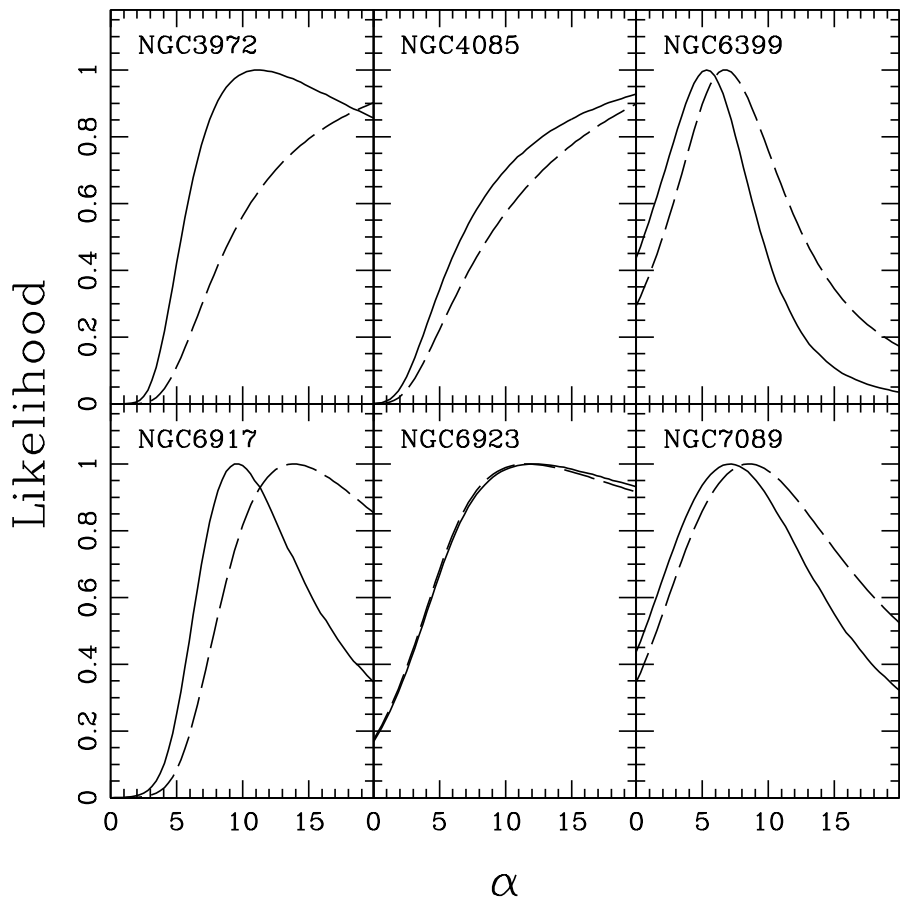}
\end{center}
\caption{Marginalised likelihood distributions with respect to the
TeVeS $\alpha$ parameter for all six disk galaxies. The solid (dashed)
lines correspond to the Chabrier (Salpeter) Initial Mass Function,
respectively.}
\label{fig:likeli}
\end{figure}

The total baryonic content is the sum of the stellar and the gas
mass. Within the TeVeS framework, we assume that the baryonic masses
are the total masses of galaxies and we combine them with the observed
rotation curves to constrain the $\alpha$ parameter that characterises
TeVeS. The values of the likelihood probability distribution function
fits are plotted against the value of the $\alpha$ parameter. The
results are given in Fig.~\ref{fig:chi2maps} as a two-dimensional (2D)
map with $\alpha$ vs. core radius. The contours are shown at the 75\%,
90\% and 95\% confidence level for the Chabrier (red) and the Salpeter
(blue) IMF. We also show the marginalised 1D likelihood function with
respect to $\alpha$ in Fig.~\ref{fig:likeli}, with the results for the
Chabrier and the Salpeter IMF given as a solid and dashed line,
respectively.  The values for the lowest $\chi^2$ and the
corresponding values of $\alpha$ are given in Table~II, where it can
be seen that each of the galaxies analysed here gives a reasonable fit
to the data once the $\chi^2$ is minimised.  The results show that the
best fit values for the $\alpha$ parameter are in the range $\sim
(5-10)$.

\begin{table}
\begin{center}
  \begin{tabular}{l||c|ccc|ccc|}
  		&	&\multicolumn{3}{c}{Salpeter}		&\multicolumn{3}{c}{Chabrier}	\\
    \scriptsize{\textbf{Galaxy}}  & dof &$\alpha$ & $\chi^2_r$ 	& $\chi^2_r(\alpha=0)$ 	&$\alpha$ & $\chi^2_r$	& $\chi^2_r(\alpha=0)$\\
    \hline
    \scriptsize{NGC 3972} &  $9$ & $\rightarrow\infty$ & $5.20$ & $38.05$ &$11.19$ &$2.47$  &$28.13$\\
    \scriptsize{NGC 4085} &  $5$ & $\rightarrow\infty$ & $3.25$ & $21.33$ &$\rightarrow\infty$	&$0.63$	&$15.45$\\
    \scriptsize{UGC 6399} &  $7$ &$6.73$  & $1.06$  &  $3.51$ 	&$5.32$	 &$1.15$ &$2.81$\\
    \scriptsize{UGC 6917} &  $9$ &$13.84$  & $1.12$  & $20.91$ 	&$9.56$	 &$1.23$ &$17.59$\\
    \scriptsize{UGC 6923} &  $4$ &$11.66$  & $1.31$  &  $4.8$ 	&$12.14$ &$1.30$ &$4.87$\\
    \scriptsize{UGC 7089} & $10$ &$8.53$   & $0.58$  &  $2.71$ 	&$7.12$	 &$0.6$ &$2.26$	\\
    \hline
  \end{tabular}
  \label{tab:chifit}
  \caption{Details of the TeVeS parameter fits for the rotation
    curves, given for both choices of the Initial Mass Function:
    Salpeter and Chabrier . Note, that the degrees of freedom are
    denoted by d.o.f. When the best fit parameter is
    $\alpha\rightarrow\infty$, it represents cases where the best fit
    is given by GR (TeVeS tends to GR as $\alpha$ goes to
    infinity). In this paper we also discuss the $\alpha=0$ case
    (which gives the lowest Dark Matter fractions for strong lensing
    in TeVeS. For comparison we include the reduced $\chi^2$ for this
    case.}
\end{center}
\end{table}

\begin{figure}[h]
\begin{center}
\includegraphics[width=8.5cm]{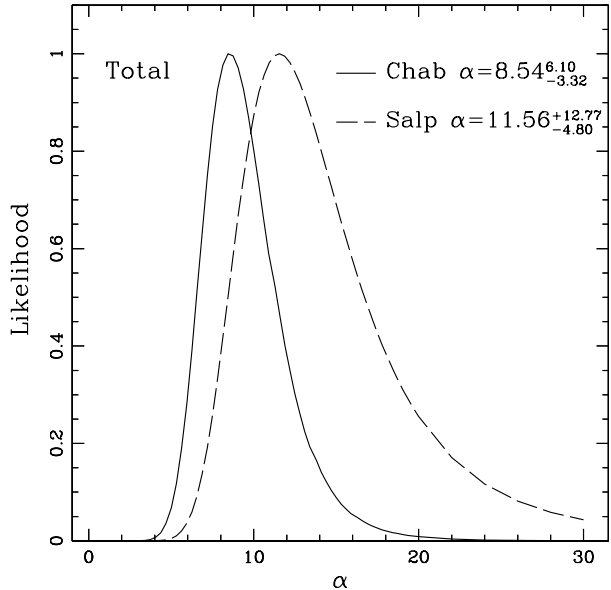}
\end{center}
\caption{Total marginalised likelihood distributions with respect to the
TeVeS $\alpha$ parameter combining the data for all six disk galaxies. The
result for the Chabrier (Salpeter) IMF is shown as a solid (dashed) line.}
\label{fig:likelitot}
\end{figure}

Figure~\ref{fig:likelitot} shows the result for the combined
probability distribution function over the six galaxies. By
normalising the total probability distribution function and setting
confidence limits at the 95\% level, the uncertainty for the best fit
value is calculated: $\alpha = 11.56_{-4.80}^{+12.77}$ for the
Salpeter IMF and $\alpha = 8.54_{-3.32}^{6.10}$ for the Chabrier
IMF. The constraints on the $\alpha$ parameter are then used in an
analysis of strong gravitational lensing on galaxy scales in TeVeS, to
assess the compatibility of these two different ways to constrain mass
over similar length-scales.

\section{Gravitational Lensing in TeVeS\label{lensing}}

In this section we perform a gravitational lensing analysis of the
TeVeS models. Given the form of our metric, Eq.(\ref{metsys}), we can
derive the equation for the deflection of light in the physical
metric. It reads~\cite{msy}
\bea
\Delta\phi=2\int^\infty_{r_0}\frac{1}{r}
\left[e^{\tilde{\zeta}(r)-\tilde{\nu}(r)}\frac{r^2}{b^2}-1\right]^{-1/2}\
{\rm d}r -\pi~,
\label{def}
\eea
where $b$ is the distance of closest approach for the incoming light ray
and it is related to $r_0$, the impact parameter through
\bea
b^2=e^{\tilde{\zeta}(r_0)-\tilde{\nu}(r_0)}r_0^2~.
\eea
Once more, these equations can be numerically solved using the same
definitions of the TeVeS constants given earlier. In order to perform
the lensing analysis in TeVeS, we use the result, Eq.~(\ref{def}), for
the deflection angle in the lensing equation (see \emph{e.g.} \cite{deflec})
\bea
\beta = \theta - \alpha(\theta,M,b)\frac{D_{\rm ls}}{D_{\rm s}}~,
\label{eq:lens}
\eea which relates the actual position of the background source
$\beta$ to the observed position of the images, given by $\theta$. For
a given cosmological model, the angular diameter distance from the
lens to the source, $D_{{\rm l}s}$, and from the observer to the
source, $D_{s}$, are both taken from the redshifts. The lensing
equation is applied independently to the multiple images of the
background source in order to find the actual solution of the lensing
model, which gives the actual position of the source ($\beta$) and the
mass of the lens.  Given that these lenses are early-type galaxies, a
Hernquist profile~\cite{hern} is adopted.  This profile has been
chosen because its projection gives the characteristic $R^{1/4}$
surface brightness profile of this type of galaxies (see
Ref.~\cite{fsy} for details). We adopt a concordance cosmology, namely
$(\Omega_{\rm m},\Omega_\Lambda,\Omega_{\rm k})=(0.3,0.7,0)$.  As
shown in Ref.~\cite{msy}, deviations from this cosmological picture
lead to very little difference in the analysis.

\begin{figure}[h]
\begin{center}
\includegraphics[width=12cm]{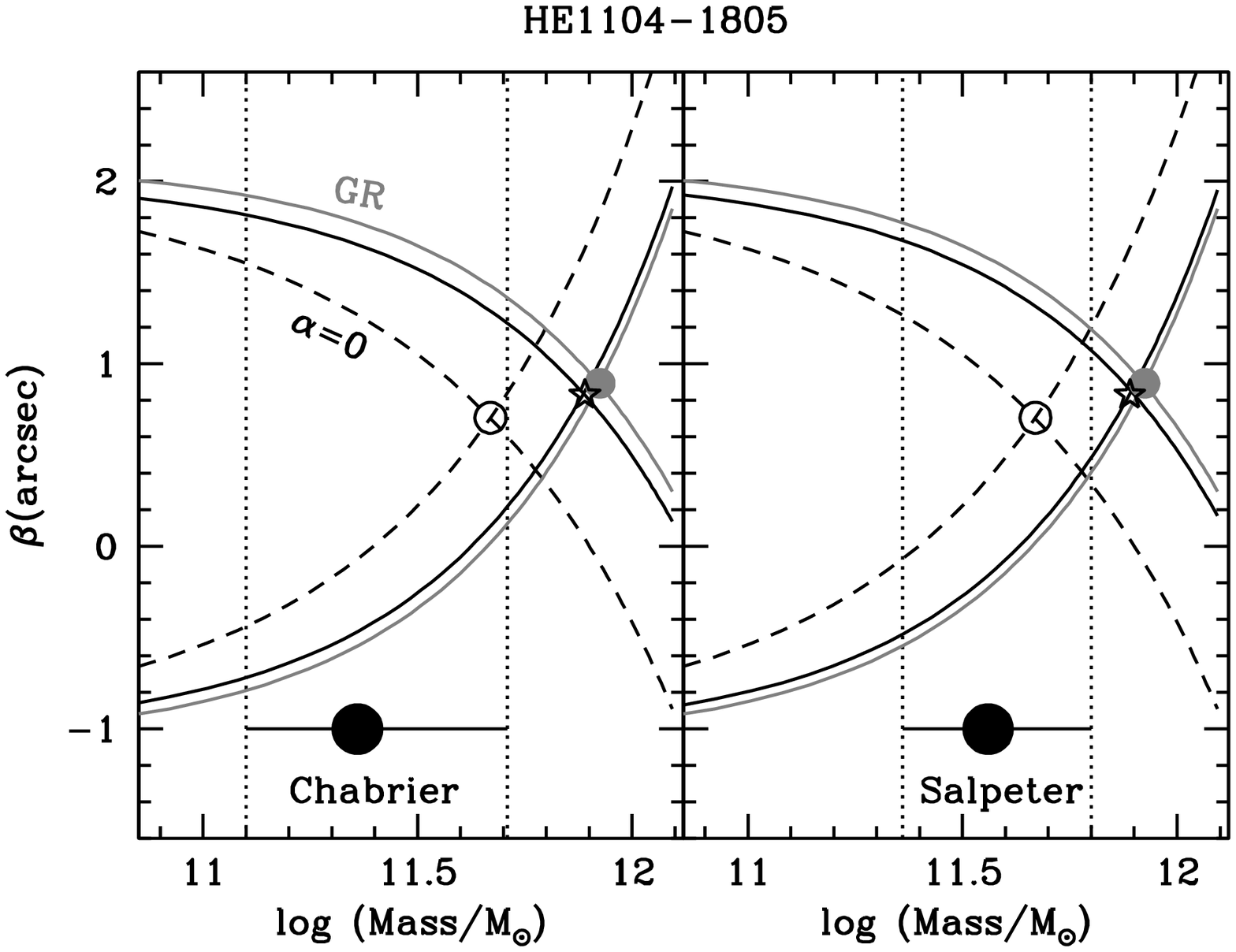}
\end{center}
\caption{Constraining the lensing mass of galaxy HE1104-1805
  (z$=0.73$).  The lines come in pairs, and represent the constraint
  on true source position ($\beta$, vertical axis) and lensing mass
  (horizontal axis) from each of the two observed positions of the
  images of the background source. The intersection of each pair of
  lines give the lensing mass. The standard GR solution is shown as a
  grey solid line, the best fit is shown as a black solid line and the
  $\alpha=0$ case discussed in the text is given as a dashed black
  line. Each panel corresponds to a different choice of Initial Mass
  Function, as labelled. The stellar mass is given as a large black
  dot along with error bars, from Ref.~\cite{fsw05}.}
\label{fig:lens}
\end{figure}

Figure~\ref{fig:lens} illustrates the methodology we use to determine
the lensing masses, for lens HE1104-1805 as an example. The lensing
equation, Eq.~(\ref{eq:lens}) is represented by the pairs of curved lines
that intersect at the true value of the lens position and lensing
mass. Three pairs of lines are shown for the best fit case of TeVeS
(solid black); standard GR (grey solid) and the $\alpha=0$ case
(dashed), which would be the best choice for a lensing based study,
although as shown above, it is significantly ruled out by the rotation
curve data. The figure shows the result both for a Chabrier ({\sl
  left}) and a Salpeter IMF, with the stellar masses given by the big
solid dot and the error bars.

To calculate a possible dark matter component, a sample of double
lensing systems from the CASTLES database~\footnote{\tt
  http://cfa-www.har vard.edu/glensdata} is analysed and the mass of
the lensing galaxy in GR and TeVeS is calculated, using the same fully
relativistic method outlined in Ref.~\cite{msy}. By comparing the mass
from lensing to the stellar mass content calculated from a comparison
of photometry and stellar population synthesis using a Chabrier
Initial Mass Function (IMF), as in Ref.~\cite{fsw05}, the mass deficit
which belonged to the ``dark'' sector is found.

From the rotation curve analysis, the favoured values of $\alpha$ are
$11.56$ and $8.54$ for Salpeter and Chabrier IMFs, respectively, and
these values are examined in the context of lensing. The results from
the lensing analysis are given in Table~III.

\begin{table}
\label{tab:lens}
\begin{tabular}{l||cc|ccc|cc|cc|ccc}
  \hline
  				&			&			& 		&$Chab$		&$Sal$		&\multicolumn{2}{c}{$\alpha = 11.56_{-4.80}^{+12.77}~Sal$}	&\multicolumn{2}{c}{$\alpha = 8.54_{-6.1}^{+3.32}~Chab$}	&{$\alpha = 0$}	&$Chab$	&$Sal$	\\
  \scriptsize{\textbf{Lens}}	&$M_{*}^{Chab}$		&$M_{*}^{Sal}$		&$M_{GR}$	&$\%DM$	&$\%DM$	&$M_{TeV}$			&$\%DM$				&$M_{TeV}$			&$\%DM$				&$M_{TeV}$	&$\%DM$	&$\%DM$	\\
  \hline
  \scriptsize{HS0818+1227} 	&$16.2^{21.2}_{13.0}$	&$20.8^{28.1}_{13.4}$	&$32.71$	&$50.5$ 	&$36.4$		&$30.59_{-0.92}^{+0.81}$	&$30.0_{-2.1}^{+1.8}$		&$30.11_{-1.00}^{+0.79}$	&$46.2_{-1.9}^{+1.4}$		&$19.1$		&$15.2$		&$-8.9$		\\
  \scriptsize{FBQ0951+2635} 	&$1.1^{2.1}_{0.5}$ 	&$1.5^{3.0}_{0.8}$	&$2.98$	  	&$63.1$ 	&$49.7$		&$2.84_{-0.06}^{+0.05}$		&$47.2_{-1.2}^{+0.9}$		&$2.81_{-0.07}^{+0.05}$		&$60.9_{-1.0}^{+0.6}$		&$1.9$		&$42.1$		&$21.0$	 	\\
  \scriptsize{HE1104-1805} 	&$22.8^{51.2}_{12.7}$ 	&$36.6^{63.7}_{23.1}$	&$83.61$  	&$72.7$ 	&$56.2$		&$78.18_{-2.36}^{+2.10}$	&$53.2_{-1.5}^{+1.2}$		&$76.96_{-2.60}^{+2.02}$	&$70.4_{-1.1}^{+0.7}$		&$47.1$		&$51.6$		&$22.3$		\\
  \scriptsize{LBQS1009-0252}	&$5.5^{7.9}_{4.2}$ 	&$7.4^{9.8}_{5.0}$	&$15.60$  	&$64.7$ 	&$52.6$		&$14.62_{-0.41}^{+0.98}$	&$49.4_{-1.5}^{+3.2}$		&$14.40_{-0.45}^{+0.36}$	&$61.8_{-1.2}^{+0.9}$		&$8.7$		&$36.8$		&$14.9$		\\
  \scriptsize{HE2149-2745}	&$4.6^{6.7}_{3.6}$	&$6.9^{8.9}_{5.0}$	&$12.03$  	&$61.8$ 	&$42.6$		&$11.27_{-0.35}^{+0.29}$	&$38.8_{-2.0}^{+1.1}$		&$11.10_{-0.35}^{+0.28}$	&$58.6_{-1.4}^{+1.0}$		&$6.3$		&$27.0$		&$-9.5$		\\
  \hline
\end{tabular}
\caption{$M_{*}$ is the stellar content of the lenses taken from
  \cite{FSans}. The other columns show the masses of the lensing
  galaxies for GR and TeVeS when $\alpha = 11.56$, $\alpha = 8.54$ and
  $\alpha = 0$. It is also given is the percentage of dark matter
  needed to account for the lensing observed given the luminous
  content of the galaxies. All masses are given in units of
  $10^{10}M_{\odot}$. The errors for the $\alpha = 11.56$ and $\alpha =
  8.54$ are derived from the errors on the value of $\alpha$ and are
  given at the 95\% confidence level.}
\end{table}

From Table~III one can clearly conclude that when $\alpha = 11.56$ and
$\alpha = 8.54$, there is a considerable need for dark matter,
comparable to that required in the pure general relativistic case. We
find that on average over the six galaxies, in the Salpeter case,
TeVeS only predicts $3.4\%$ less dark matter than the general
relativistic case: GR requires $47.5\%$ dark matter as opposed to
TeVeS which requires $44.1\%$. For the Chabrier case TeVeS gives on average
$59.6\%$ dark matter as opposed to GR which gives $62.6\%$, a reduction of only $5.0\%$. Using the errors on the value of $\alpha$, the
errors on the mass are calculated at the $95\%$ confidence level. The
error bars are quite tight, only $3.3\%$ on average in the Salpeter
case and $2.2\%$ for Chabrier, showing just how strongly the analysis
rules out this class of $\mu(y)$ functions. This result is in
agreement with the work detailed in ref.~\cite{msy}, where it was
shown that even for a value of $\alpha$ as low as 0 there is a
\emph{not-insignificant} need for dark matter. However the analysis
presented in Ref.~\cite{msy} shows that only values of $\alpha$
greater than or equal to 1 can be definitively ruled out by the
lensing data due to the error bars on the stellar masses. The $\alpha
= 0$ case is not ruled out as the lensing masses are within the error
bars of the stellar masses, although they are consistently just within
the upper error bar which suggests that with higher accuracy stellar
mass estimates, even this case could be ruled out using just a lensing
analysis. In the absence of more accurate estimates, we can test the
$\alpha = 0$ with rotation curves. Table~II comparison of the $\chi^2$
for rotation curves for different values of $\alpha$. As it can be
clearly seen, for three of our sample of six galaxies (N3972, N4085
and U6917), the $\alpha = 0$ gives a value of $\chi^2$ which is nearly
an order of magnitude larger than the best fit $\alpha$ case for both
Salpeter and Chabrier IMFs, and is sufficiently high so as to be
considered a poor fit. This implies that the $\alpha = 0$ case is
\emph{incompatible} with rotation curves. This result was also
suggested in Refs.~\cite{mu1, mu2} where the authors showed, using a
non-relativistic photometric fitting approach to rotation curves as
opposed to the relativistic parametric fitting used here, that galaxy
NGC3198 could not fit the data when $\alpha = 0$ but only when $\alpha
= 1$, a case already ruled out by the lensing analysis of
Ref.~\cite{msy}.
  
From this we can infer that there is \emph{ no value} of $\alpha$
which can satisfactorily fit \emph{both} rotation curves and
gravitational lensing.  In conclusion, at the very least an entirely
different form of the $\mu(y)$ function, and its related MONDian
equivalent $f(x)$, needs to be found if the modifications to gravity
are to remain universal in applicability with only the acceleration
scale being free to be fitted.

\section{Conclusions and Outlook\label{concl}}

In this paper we have made an attempt to constrain TeVeS models by
using both gravitational lensing and rotation curve data. In
particular, we have analysed the one-parameter models that have been
fitted against the rotation curves, and we have found that there is no
value of the parameter $\alpha$ that fits both the rotation curves and
the gravitational lensing data of galaxies. The baryonic mass of the
galaxies is calculated using photometric data and is assumed to
account for the total mass budget of the system within the TeVeS
paradigm. A standard likelihood method gives $\alpha =
11.56_{-4.80}^{+12.77}$ for the Salpeter IMF, and $\alpha =
8.54_{-6.1}^{+3.32}$ for the Chabrier IMF, at a 95\% confidence level.
Consequently, we estimate the mass content of five strong
gravitational lenses from the CASTLES survey and compare their lensing
masses to the corresponding stellar content, calculated from
photometry.  On taking into account the constraint from rotation
curves, we find that the lensing mass within the TeVeS formalism still
shows an excess around 50$\%$ over the baryonic content. The only
successful parameter value from lensing ($\alpha = 0$) is shown to be
incompatible with rotation curves.

For the fits we used a particular set of galaxies, for which
information for the inner part of the velocity distribution is
available. The upshot of our analysis is that, at least in its
simplest original form, which by the way is the only one studied in
the literature so far, TeVeS is ruled out. One may think that
multi-parametric extensions of TeVeS models may be
successful. However, such models have not been proposed so
far. Lacking a fundamental microscopic derivation of the
$\mu$-functions, the use of such complicated models becomes unnatural
and probably of no physical relevance.

Several features of the TeVeS model cannot be excluded at present, and
in fact such features may be desirable and essential for fitting
extended versions of TeVeS which include dark matter components.  In
particular, the bi-metric nature of TeVeS is an important ingredient
which may indeed characterise more fundamental models, such as
string-inspired foam ones~\cite{nickmairi}. Moreover, the vector-like
cosmological instabilities implied by the vector time-like field
$U^\mu$ may play an important r\^ole in structure formation, as
emphasised in Ref.~\cite{liguori}, reconciling the observed baryonic
spectrum with theoretical predictions.  It is worthy of mentioning at
this stage that in the context of the string models of
Ref.~\cite{nickmairi}, the bi-metric feature of TeVeS is realised
because of the presence of non-trivial dilaton fields, while the
vector instabilities arise because of physical interactions of open
string matter with space-time defects that such models include, as a
result of the recoil of the latter. The models have a non-trivial dark
energy component built in, as a result of the Born-Infeld-type
dynamics of the vector field, that is characteristic in open string
theories on brane worlds considered in that work.  In addition, the
string spectrum contains dark matter particles, in particular
supersymmetric ones, and in this sense the TeVeS-like features
co-exist with extra components of dark matter and dark energy in the
model, fitting perfectly all currently available data, including
cosmological ones.  We should stress, however, that in such string
models, the fundamental physics is entirely different, and even if one
has TeVeS-like features, such as bi-metric models and vector-like
instabilities, these are features that pre-existed the specific TeVeS
models of Ref.~\cite{bek} and their origin is traced back to
fundamental structures in string/brane framework.

Also, for reasons explained in Ref.~\cite{nickmairi}, an important
r\^ole is played by neutrino matter in such models, which provide a
significant component of dark matter, in addition to that offered by
the supersymmetric matter. In this latter respect, we also mention
other phenomenological works within the context of the TeVeS model or
extensions thereof~\cite{sko06,neutrinos}, in a conventional field
theoretic framework, claiming a prominent r\^ole of neutrinos as dark
matter components in TeVeS-like models, fitting cosmological
data. These are interesting avenues for research, which we intend and
hope to be able to pursue in the near future.

\section*{Acknowledgements}
The work of N.\,E.\,M. and M.\,S. is
partially supported by the European Union through the Marie Curie
Research and Training Network UniverseNet (MRTN-CN-2006-035863), while
that of M.F.Y. is supported by an E.P.S.R.C. studentship.


\end{document}